\documentclass{aip-cp}

\usepackage[numbers]{natbib}
\usepackage{rotating}
\usepackage{graphicx}

\begin{document}

% Title portion
\title{Beta Decays in Investigations and Searches for Rare Effects}

\author[aff1]{V.I. Tretyak\corref{cor1}}

\affil[aff1]{Institute for Nuclear Research, 03028 Kyiv, Ukraine}

\corresp[cor1]{Corresponding author: tretyak@kinr.kiev.ua}

\maketitle

\begin{abstract}
Current status of experimental investigations of rare single $\beta$ decays ($^{48}$Ca, $^{50}$V, $^{96}$Zr, $^{113}$Cd, $^{113m}$Cd, $^{115}$In, $^{123}$Te, $^{180m}$Ta, $^{222}$Rn) is reviewed. Nuclei which decay through single $\beta$ decay very often constitute backgrounds in studies of rare effects like double beta decay, solar neutrinos or dark matter. Summary of correction factors used in description of forbidden $\beta$ decays is also briefly given. 
\end{abstract}

\section{INTRODUCTION}

Beta radiation was observed long ago \cite{Rut99} but our knowledge of this phenomenon still can be and should be improved.
Some rare $\beta$ decays ($T_{1/2} > 10^{10}$ y) are poorly investigated (spectrum shape is not measured, e.g. $^{50}$V) and even not observed (e.g. $^{123}$Te, $^{180m}$Ta).
Interest to $\beta$ decays increased during last time because sometimes they create significant backgrounds in searches for and investigations of rare effects, like solar neutrinos (e.g. $^{14}$C in Borexino \cite{Bel14a}), double beta ($2\beta$) decay (e.g. $^{39}$Ar, $^{42}$Ar/$^{42}$K in GERDA \cite{Ago17}) or dark matter experiments, especially based on Ar (e.g. $^{39}$Ar, $^{42}$Ar/$^{42}$K in DarkSide \cite{Agn16}). While in the above-mentioned cases $\beta$ decays constitute one of the main features in the measured spectra, other single $\beta$ decayers create not so big but noticeable backgrounds in many experiments: $^{40}$K, $^{90}$Sr/$^{90}$Y, $^{137}$Cs, $^{214}$Bi and others. Quite often their energy spectra have not allowed shape but are classified as forbidden (unique or non-unique) of some level of forbiddenness, and sometimes they are not well studied. For example, $^{214}$Bi is one of the main backgrounds in all $2\beta$ experiments due to its high energy release, $Q_\beta = 3270$ keV. In 19.1\% it decays to the ground state of $^{214}$Po with change in spin and parity $1^- \to 0^+$, $\Delta J^{\Delta \pi}=1^-$, classified as 1-forbidden non-unique (1 FNU); however, to our knowledge, theoretical calculations of its shape are absent in the literature; also, it was not well measured experimentally\footnote{Its shape in graphical form can be found only in old papers (see \cite{Ber52,Kag53,Ric55}), from which it could be concluded that it is not far from the allowed.}. 
It is clear that good knowledge of shapes of single $\beta$ decays is very important for a proper fitting of experimental spectra and correct estimation of little effects possibly present in these spectra. 

Below, after a summary on classification of single $\beta$ decays and shapes of their energy spectra, we review recent achievements in studies and searches for rare $\beta$ processes.  

\section{SHAPES OF BETA SPECTRA}

Beta decays are classified as allowed or forbidden of some level of forbidenness in dependence on change in spin $J$ and parity $\pi$ between mother and daughter nuclei: 

\begin{table}[htb]
\begin{tabular}{llll}
$\Delta J^{\Delta \pi}=$ & $0^+, 1^+$ & ~  & –- allowed; \\
~                        & $0^-, 1^-, 2^+, 3^-, 4^+, ...$ & $\Delta \pi = (-1)^{\Delta J}$ & -– forbidden non-unique (FNU); forbidenness = $\Delta J$; \\
~                        & ~~~~~~~~~~ $2^-, 3^+, 4^-, ...$ & $\Delta \pi = (-1)^{\Delta J-1}$ & -– forbidden unique (FU); forbidenness = $\Delta J-1$. \\
\end{tabular}
\end{table}

For unique decays, rate of decay and shape of spectrum are defined by only one nuclear matrix element.
Shape of $\beta$ spectrum in general is described as:
$\rho(E) = \rho_{allowed}(E) \times C(E)$, 
where 
$\rho_{allowed}(E) = F(Z,E)WP(Q_\beta-E)^2$ is the distribution for the allowed spectrum;
$W(P)$ is the total energy (momentum) of $\beta$ particle;
$F(Z,E)$ is the Fermi function:
$F(Z,E)=const \cdot P^{2\gamma-2}\exp(\pi s)\mid\Gamma(\gamma+is)\mid^2$,
where $\gamma=\sqrt{1-(\alpha Z)^2}$, $s=\alpha ZW/P$, $\alpha=1/137.036$ is the fine structure constant, $Z$ is the atomic number of the daughter nucleus ($Z>0$ for $\beta^-$ and $Z<0$ for $\beta^+$ decay), and $\Gamma$ is the gamma function;
$C$ is the (empirical) correction factor; 	
$W$ is in $m_ec^2$ units and $P,Q$ below -- in $m_ec$ units.

For FNU	decays correction factors very often have the following general forms:
$C_1(E) = 1+a_1/W+a_2W+a_3W^2+a_4W^3$ or
$C_1(E) = 1+b_1P^2+b_2Q^2$, where $Q$ is the momentum of (anti)neutrino.

For FU decays correction factors often are given as: 
$C = C_1C_2$, where $C_2$ is:
1 FU --	$C_2 = P^2+c_1Q^2$;
2 FU --	$C_2 = P^4+c_1P^2Q^2+c_2Q^4$;
3 FU -- $C_2 = P^6+c_1P^4Q^2+c_2P^2Q^4+c_3Q^6$;
4 FU --	$C_2 = P^8+c_1P^6Q^2+c_2P^4Q^4+c_3P^2Q^6+c_4Q^8$.
Sometimes alternative forms are used: 
1 FU --	$C_2 = Q^2+\lambda_2P^2$, 2 FU -- analogous expression with $\lambda_2, \lambda_4$, and so on, where $\lambda_i$ are the Coulomb functions calculated in \cite{Beh69}.

Coefficients $a_i, b_i, c_i$ above should be calculated theoretically (they are mixture of products of phase space factors with different nuclear matrix elements) or extracted from experimental measurements. Compilations of the experimental $a_i, b_i, c_i$ can be found in \cite{Pau66,Dan68,Beh76,Mou16}.

As examples, Fig. 1 (left pannel) shows $\beta$ spectra for $^{39}$Ar and $^{42}$Ar in comparison with the allowed shapes. In both cases $\Delta J^{\Delta \pi}=2^-$, so decays are classified as 1~FU; correction factor is given as
$C(E) = Q^2+\lambda_2P^2$. 

The middle panel of Fig.~1 shows $\beta$ spectra of $^{42}$K (daughter of long-living $^{42}$Ar), measured in few works. For transition $^{42}$K $\to$ $^{42}$Ca (ground state, probability 81.90\%), which is also 1~FU, correction factor is 
$C(E) = (Q^2+\lambda_2P^2)(1+aW)$. 
Transition $^{42}$K $\to$ $^{42}$Ca (excited level with $E_{exc}=1525$ keV, 17.64\%) is classified as 1~FNU ($\Delta J^{\Delta \pi}=0^-$), and 
$C(E) = 1+a_1/W+a_2W+a_3W^2$. Values of $a, a_i$ and references to original works can be found in \cite{Beh76}.

The right pannel shows spectra of $^{40}$K and $^{137}$Cs. For $^{40}$K $\to$ $^{40}$Ca (g.s., 89.28\%,  $\Delta J^{\Delta \pi}=4^-$, 3~FU) corection factor is
$C(E) = P^6+c_1P^4Q^2+c_2P^2Q^4+c_3Q^6$ with $c_1=c_2=1, c_3=7$ measured in \cite{Kel59}.
For $^{137}$Cs $\to$ $^{137}$Ba (g.s., 5.3\%,  $\Delta J^{\Delta \pi}=2^+$, 2~FNU) corection factor is
$C(E) = 1+a_1/W+a_2W+a_3W^2$ with $a_1=0, a_2=-0.6060315, a_3=0.0921520$ measured in \cite{Hsu66}.
	
This is a demonstration that forbidden $\beta$ spectra can significantly deviate from the allowed shapes, and it is necessary to take this into account in simulations of corresponding backgrounds. It should be also noted that sometimes even decays which are classified as allowed have deviations from the allowed shape, e.g. $^{14}$C, where in theoretical description the first order terms mutually compensate each other and the second order terms fetermine the $T_{1/2}$ and the shape. 

\begin{figure}[h]
\centerline{\includegraphics[width=150pt]{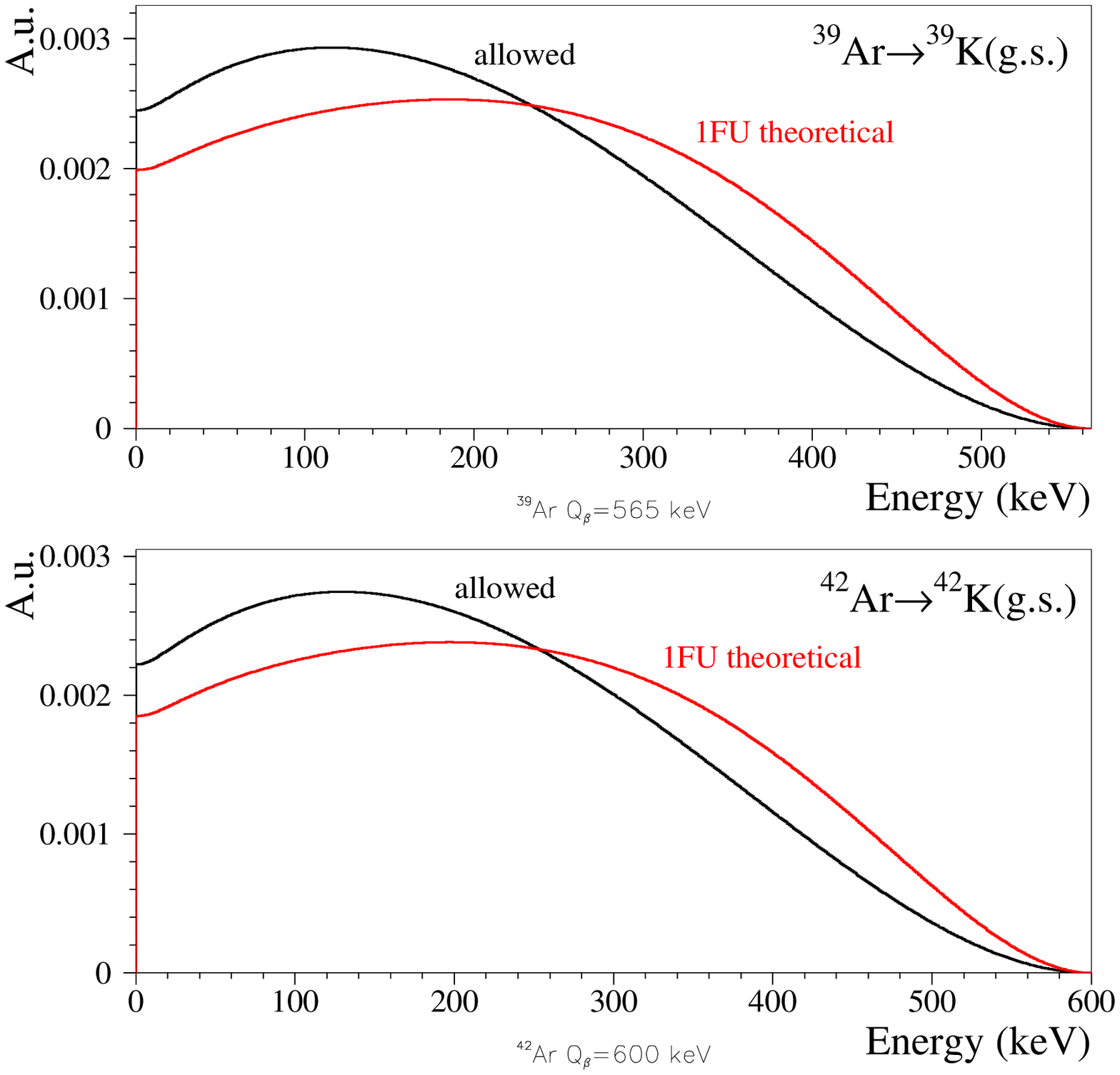}
~\includegraphics[width=150pt]{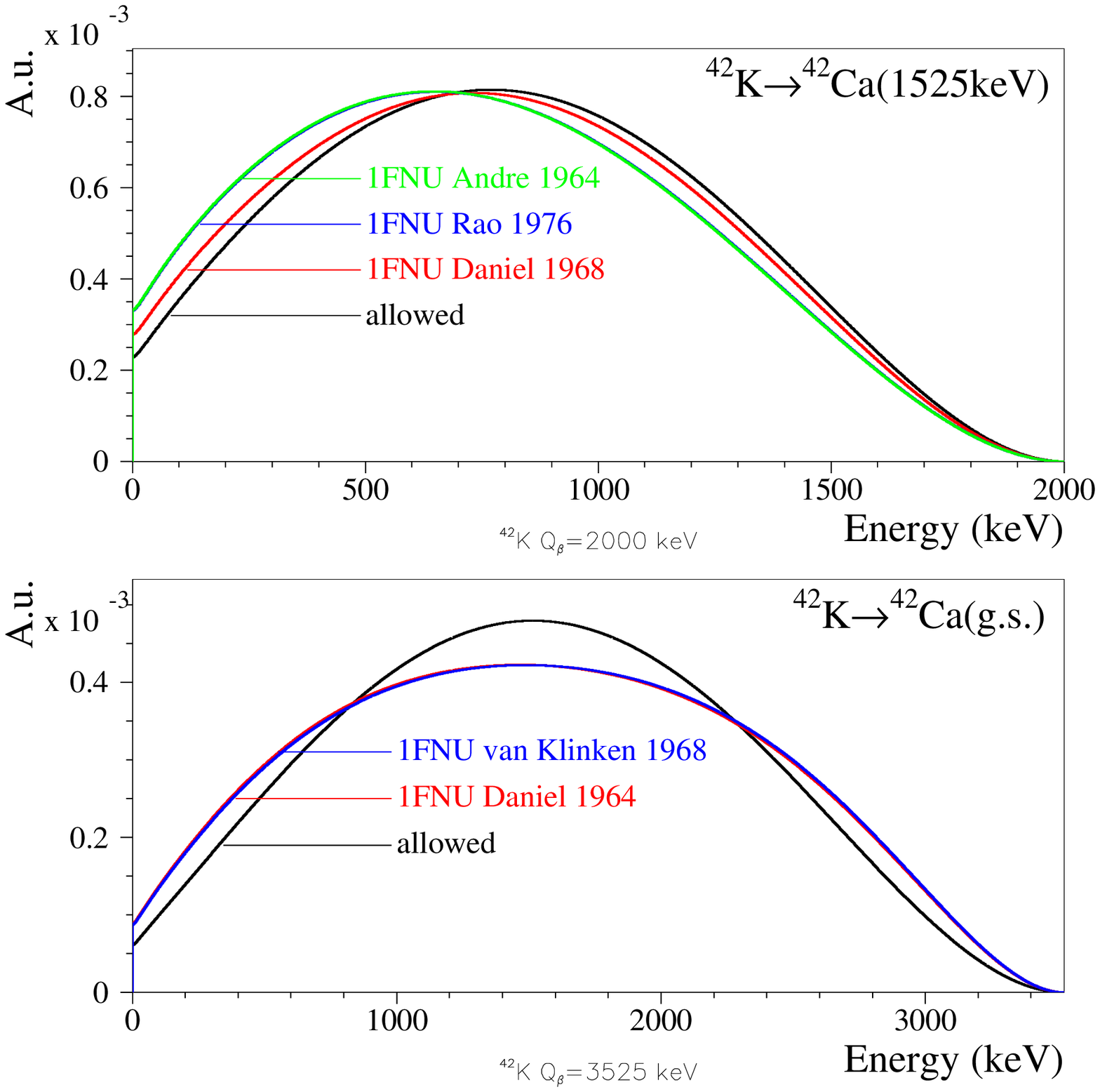}
~\includegraphics[width=150pt]{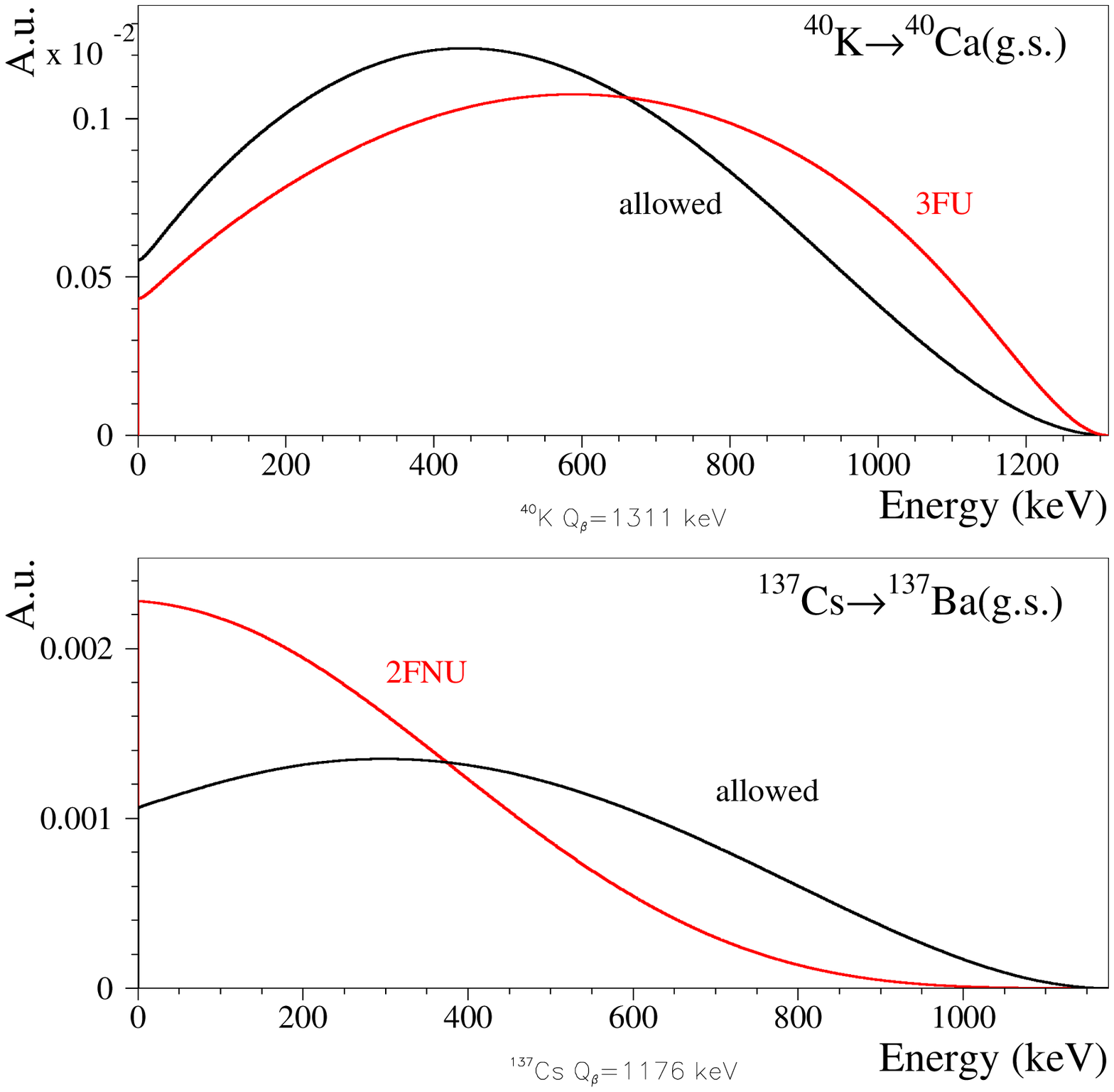}}
\caption{Shapes of $\beta$ spectra of $^{39,42}$Ar, $^{42}$K, $^{40}$K and $^{137}$Cs.}
\end{figure}

\section{INVESTIGATIONS OF RARE BETA DECAYS}

\fbox{\textbf{$^{48}$Ca}.}
$^{48}$Ca can decay through single $\beta$ decay populating few levels in $^{48}$Sc, however, transition to the 131 keV level is estimated as the most probable with $T_{1/2}=(2.6-7.0)\times10^{20}$ y \cite{Haa14}. 
The process is not observed yet, and the best experimental limit $T_{1/2}>2.5\times10^{20}$ y \cite{Bak02} is not far from the above theoretical estimations. 
It should be noted that the second order process -- two neutrino $2\beta$ decay to $^{48}$Ti -- is faster in this case; it is already observed with 
$T_{1/2}=6.4\times10^{19}$ y \cite{Arn16}.
\\
\fbox{\textbf{$^{50}$V}.} $^{50}$V is one of only 3 nuclei where $\beta$ processes with $\Delta J^{\Delta \pi}=4^+$ were observed (other two are $^{113}$Cd and $^{115}$In). Low natural abundance ($\delta=0.250\%$) and big $T_{1/2}$ make its investigations difficult. In the most sensitive to-date experiment \cite{Dom11}, only limit for $\beta$ decay to $^{50}$Cr was found as: 
$T_{1/2}>1.7\times10^{18}$ y (not confirming some earlier observations), 
while electron capture (EC) to $^{50}$Ti was measured with 
$T_{1/2}=(2.3\pm0.3)\times10^{17}$ y.
\\
\fbox{\textbf{$^{96}$Zr}.} Situation with $^{96}$Zr is analogous to $^{48}$Ca: single $\beta$ decay to $^{96}$Nb is possible but not yet observed ($T_{1/2}>3.8\times10^{19}$ y \cite{Arp94} for the most probable transition to the 44 keV excited level), while $2\beta$ decay is already measured with 
$T_{1/2}=(2.3\pm0.2)\times10^{19}$ y \cite{Nov13}.
\\
\fbox{\textbf{$^{113}$Cd}.} In the last work where the half-life and shape of $^{113}$Cd $\beta$ spectrum ($\Delta J^{\Delta \pi}=4^+$, 4~FNU) was precisely measured \cite{Bel07}, coefficients $c_i$ in correction factor 
$C(E) = P^6+c_1P^4Q^2+c_2P^2Q^4+c_3Q^6$ were determined. While the experimental spectrum was perfectly described with this $C(E)$, it is interesting to note that, in fact, this factor usually is used for decays with $\Delta J^{\Delta \pi}=4^-$ (3~FU).
Recently it was noted \cite{Haa16,Haa17,Kos17} that, because for non-unique forbidden $\beta$ decays shape of energy spectrum depends on sum of different nuclear matrix elements with different phase space factors which include also the weak interaction coupling constants $g_A$ and $g_V$, it is possible to find the $g_A$ and $g_V$ values by comparing theoretical shape with the experimental spectrum. This observation is very important for predictions of $T_{1/2}$'s for $2\beta$ decays because $T_{1/2}(2\beta) \sim g_A^4$ and known uncertainties in the $g_A$ value could result in $1-2$ orders of magnitude uncertainty in $T_{1/2}$.
\\
\fbox{\textbf{$^{113m}$Cd}.} The excited state of $^{113}$Cd ($E_{exc}=263.5$ keV) has a quite long half-life of $\simeq14$ y. It decays to the ground state of $^{113}$In ($\Delta J^{\Delta \pi}=1^-$, 1~FNU). Its shape is under measurements by the DAMA-KINR collaboration at the Gran Sasso National Laboratories with the help of $^{106}$CdWO$_4$ crystal scintillator contaminated with $^{113m}$Cd; some deviations from the allowed shape are found \cite{Bel17}.
\\
\fbox{\textbf{$^{115}$In}.} While for $^{113}$Cd shape of $\beta$ spectrum was measured in quite big number of works, shape of $^{115}$In was measured only in one old work \cite{Pfe79}. This experiment, done with liquid scintillator (LS) loaded by In at 51.2 g/L, had a number of drawbacks: measurements were performed at the sea level (thus background was quite high); quenching of low energy electrons, which is strong for LS, was not taken into account; energy resolution was not known exactly; and energy threshold was quite high (around 50 keV). Remeasuring of this decay in low background conditions would be very interesting. Such a possibility appeared recently with LiInSe$_2$ scintillating bolometer ($8\times15\times19$ mm, 10.2 g) which is under measurements now at the Modane underground laboratory \cite{Win17}.
\\
\fbox{\textbf{$^{115}$In $\to$ $^{115}$Sn$^*$}.} Decay of $^{115}$In to the first excited level of $^{115}$Sn ($E_{exc}=497.334(22)$ keV) was at the first time observed in \cite{Cat05} and further confirmed in \cite{Wie09,And11}. Precise measurements of the atomic mass difference between $^{115}$In and $^{115}$Sn, $\Delta M_A = 497.489\pm0.010$ keV \cite{Mou09}, allowed to conclude that $^{115}$In $\to$ $^{115}$Sn$^*$ is the $\beta$ decay with the lowest known $Q_\beta^*$ value of only $155\pm24$ eV. Very recently also the energy of the excited $^{115}$Sn level was remeasured more precisely as:
497.316(7) keV \cite{Urb16} (that results in $Q_\beta^*=173\pm12$ eV) and 497.341(3) keV \cite{Zhe17} ($Q_\beta^*=148\pm10$ eV). 
\\
\fbox{\textbf{$^{123}$Te}.} Electron capture of $^{123}$Te was registered in old work \cite{Wat62} with $T_{1/2}=(1.24\pm0.10)\times10^{13}$ y. However, in \cite{Ale96} it was found that the real value is 6 orders of magnitude higher: $T_{1/2}=(2.4\pm0.9)\times10^{19}$ y. Later, also this result was found incorrect, and only limit was set as $T_{1/2}>5.0\times10^{19}$ y \cite{Ale03}. Observation of \cite{Ale96} was explained by the electron capture in $^{121}$Te; this unstable isotope was created in TeO$_2$ crystals used in the measurements through neutron capture by $^{120}$Te while the crystals were at the Earth level. Natural abundance of $^{120}$Te is very small, $\delta=0.09\%$, and this is a good demonstration how tiny effect can mimick another rare effect.
\\
\fbox{\textbf{$^{180m}$Ta}.} It is interesting to note that in the natural mixture of elements $^{180}$Ta is present ($\delta=0.012\%$) not in the ground state (it quickly decays with $T_{1/2}\simeq8$ h) but in an excited state ($E_{exc}=77$ keV). Its decay (through EC and $\beta^-$) is still not found; the best limits were set in the recent work \cite{Leh17} as:
$T_{1/2}($EC$)>2.0\times10^{17}$ y and 
$T_{1/2}(\beta^-)>5.8\times10^{16}$ y.
\\
\fbox{\textbf{$^{222}$Rn}.} $^{222}$Rn is known as 100\% $\alpha$ decaying ($Q_\alpha=5590$ keV, $T_{1/2}=3.82$ d). However, it was noted recently \cite{Bel14b} that single $\beta$ decay (1~FU) is also energetically possible with $Q_\beta=24\pm21$ keV. Half-life was estimated as $6.7\times10^4-2.4\times10^8$ y, in dependence on $Q_\beta$. After $\beta$ decay of $^{222}$Rn, one should observe chain of $\alpha$ and $\beta$ decays which is different from that after its $\alpha$ decay. Looking for this chain in BaF$_2$ crystal scintillator polluted by $^{226}$Ra, only the limit for $^{222}$Rn was set as $T_{1/2}(\beta)>8.0$ y \cite{Bel14b}.

\section{CONCLUSIONS}

There was a little interest in investigations of rare $\beta$ decays since 1970's. However, during the last time, development of experimental technique lead to improvement in sensitivity, and new decays were observed with
extreme characteristics (e.g. $\beta$ decay with lowest $Q_\beta$ of $\simeq 155$ eV for $^{115}$In $\to$ $^{115}$Sn$^*$).
Interest to $\beta$ shapes also is growing, in particular for nuclides which create background in rare events' searches.
Many theoretical works also appeared last time. New approach to measure the $g_A/g_V$ ratio through non-unique forbidden beta decays ($^{113}$Cd, $^{115}$In) is proposed. It could be concluded that investigations of rare $\beta$ decays start to
revive now, and we could expect new interesting theoretical works and experimental measurements.

\section{ACKNOWLEDGMENTS}

I am grateful for partial support by the National Academy of Sciences of Ukraine (grant F8-2017 related with the IDEATE International Associated Laboratory, LIA). Support from the MEDEX'2017 Organizing Committee also is warmly acknowledged.

\nocite{*}
\bibliographystyle{aipnum-cp}%

\end{document}